\documentclass[a4paper,10pt]{iopart}
\usepackage[dvips,final]{epsfig}
\usepackage{graphicx,bm,iopams}

\newcommand{\beq}{\begin{equation}}
\newcommand{\eeq}{\end{equation}}
\newcommand{\bea}[1]{\begin{subequations}\label{#1}\begin{eqnarray}}
\newcommand{\eea}{\end{eqnarray}\end{subequations}}
\newcommand{\bal}{\begin{align}}
\newcommand{\eal}{\end{align}}

\begin{document}

\title{Omnidirectional refractive devices for flexural waves based on graded phononic crystals}
\author{Daniel Torrent, Yan Pennec and Bahram Djafari-Rouhani}
\address{Institut d'Electronique, de Microl\'ectronique et de Nanotechnologie, UMR CNRS 8520, Universit\'e de Lille 
1, 59655 Villeneuve d'Ascq, France}
\ead{daniel.torrent@iemn.univ-lille1.fr}

\date{\today}

\begin{abstract}
Different omnidirectional refractive devices for flexural waves in thin plates are proposed and numerically analyzed. Their realization is explained by means phononic crystal plates, where a previously developed homogenization theory is employed for the design of graded index refractive devices. These devices consist of a circular cluster of inclusions with properly designed gradient in their radius. With this approach, the Luneburg and Maxwell lenses and a family of beam splitters for flexural waves are proposed and analyzed. Results show that these devices work properly in a broadband frequency region, being therefore an efficient approach for the design of refractive devices specially interesting for nano-scale applications. 
\end{abstract}


\maketitle


\section{Introduction}
The control of the propagation characteristics of flexural waves in thin elastic plates has been a topic of intense research. Thus, propagation properties of flexural waves in plates with periodic arrangements of rigid pins \cite{Evans2007,mcphedran2009platonic}, holes \cite{Movchan2007}, attached pillars \cite{pennec2008low,pennec2009phonon,marchal2012dynamics} or point-like spring-mass resonators \cite{Xiao2011,Xiao2012,Torrent2013Graphene} attached to them has been investigated by several groups, as well as the resonant properties of complex inclusions \cite{Zhu2012}.

These complex structures behave, in the low frequency limit, like homogeneous materials with properties that can be artificially tailored, allowing the design and fabrication of advanced refractive devices. For instance, cloaking devices \cite{Farhat2009,Farhat2009a,Stenger2012} for making objects invisible to flexural waves, as well as flat lenses based on negative refraction \cite{Farhat2010,pierre2010negative}, gradient index lenses \cite{lenteTTWu,zhao2012efficient} and omnidirectional absorbers \cite{krylov2012resumen,climente2013omnidirectional} have been recently proposed and experimentally verified. 

In a recent publication \cite{climente2014gradient} Climente et al. proposed a set of omnidirectional refractive lenses for flexural waves based on thickness variations. This is possible given that the propagation velocity of flexural waves in thin plates depends not only on the material's parameters, but also on the plate's thickness. Therefore, a gradient in this thickness will create a gradient in the refractive index and in this way several devices are possible to be built. The applications of these devices are interesting in all the scales, however the fabrication of a plate with a controlled position-dependent thickness is complicated at the nano-scale, thus an alternate approach for the fabrication of these omnidirectional devices, specially feasible at the nano-scale, is required.

In this work we propose a set of omnidirectional devices for flexural waves based on graded phononic crystal plates. In this approach, the device is consist of a circular cluster of inclusions arranged in an ordered lattice. This cluster behaves, in the low frequency limit, as a homogeneous device whose elastic parameters depend on the physical nature of the inclusions and their size \cite{torrent2014effective}. Then, if the size of the inclusions is changed according to their position, following
a specific law, the refractive index of the material can be changed accordingly. This approach is specially interesting at the nano-scale, where current fabrication techniques allow easily the creation of regular arrays of inclusions with different radii.

The paper is organized as follows. After this introduction, Section \ref{sec:theory} explains the homogenization theory employed and defines the effective refractive index for flexural waves. Section \ref{sec:LNMX} analyzes the creation of two classical omnidirectional refractive devices: Luneburg and Maxwell lenses. Section \ref{sec:BS} is devoted to the study of an interesting family of these devices, named ``beam splitters'' and, finally, Section \ref{sec:summary} summarizes the work.

\section{Effective Refractive Index for Flexural Waves}
\label{sec:theory}
The propagation of flexural waves in thin plates can be described, for wavelengths larger than the thickness
of the plate $h_b$, by means of the bi-Helmholtz equation \cite{Graff}
\beq
\label{eq:flexwaves}
(\nabla^4-k_b^4)W(x,y)=0,
\eeq
where $W(x,y)$ is the vertical displacement of the plate and the wavenumber is given by
\beq
\label{eq:wavenumber}
k_b^4=\frac{\rho_bh_b}{D_b}\omega^2,
\eeq
being $\rho_b$ the mass density of the plate, $h_b$ its thickness and $D_b$ the rigidity of the plate, related with the plate's Young modulus $E_b$ and Poisson's ratio $\nu_b$ as
\beq
D_b=\frac{E_bh_b^3}{12(1-\nu_b^2)}.
\eeq

The solution of equation \eref{eq:flexwaves} satisfy the usual refraction laws, so that the refraction of waves passing from a medium 1 to a medium 2 is defined by means of the ratio between the wavelengths, thus
\beq
\label{eq:n}
n_{12}=\frac{k_1}{k_2}=\left(\frac{\rho_2 h_2}{D_2}\frac{D_1}{\rho_1 h_1}\right)^{1/4}
\eeq
where equation \eref{eq:wavenumber} has been used. It is clear from equation \eref{eq:n} that the refractive index can be properly designed by means of the material's properties, as usual, but also by means of the plate's thickness. This last idea has been exploited by krylov et al. \cite{krylov2012resumen} and Climente et al. \cite{climente2013omnidirectional,climente2014gradient} for the creation of gradient index devices by means of thickness variations. 

From the practical point of view, the fabrication of thin plates with a region of thickness variation following a specific profile can be complex under certain conditions, specially at the micro or nano-scales. For this reason, in this work we propose an alternate though complementary way for the realization of these refractive devices based on phononic crystals.

\begin{figure}\centering
\centering
\includegraphics[scale=1]{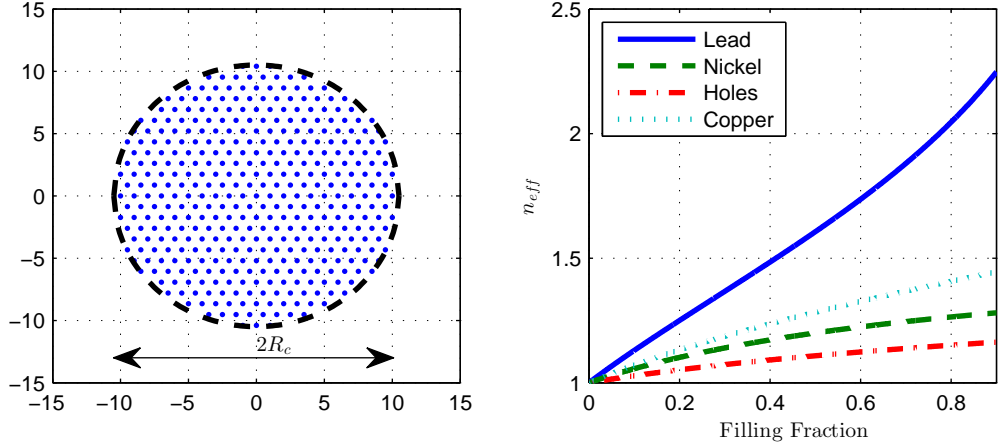}
\caption{\label{fig:schematics}
Left panel: Schematic view of the cluster of inclusions employed for the design of omnidirectional lenses
in thin plates. Right panel: Effective refractive index as a function of filling fraction for circular inclusions in a Silicon plate for different materials. As we see, lead inclusions produce the wider range for the refractive index.
				}
\end{figure}

The proposed refractive device is schematically shown in the left panel of  figure \ref{fig:schematics}. A circular cluster of inclusions of certain material are arranged in a triangular lattice. It is known that, for wavelengths larger than the lattice constant $a$, this cluster behaves as an effective
medium with certain effective parameters, which depend on the physical properties of the inclusions and their size. In reference \cite{torrent2014effective} it was demonstrated that the effective parameters for this arrangement of inclusions is obtained by means of the following equations
\numparts
\begin{eqnarray}
\rho_{eff}=(1-f)\rho_b+f\rho_a\\
\fl
D_{eff}(1+\nu_{eff})=\nonumber \\
\fl \frac{(1+\nu_b)(D_b(1-\nu_b)+D_a(1+\nu_a))-f(1-\nu_b)(D_b(1+\nu_b)-D_a(1+\nu_a))}{D_b(1-\nu_b)+D_a(1+\nu_a)-f(D_b(1+\nu_b)-D_a(1+\nu_a)}D_b\\
\fl
D_{eff}(1-\nu_{eff})=\nonumber \\
\fl \frac{(1-\nu_b)(D_b(3+\nu_b)+D_a(1-\nu_a))-f(3+\nu_b)(D_b(1-\nu_b)-D_a(1-\nu_a))}{D_b(3+\nu_b)+D_a(1-\nu_a)-f(D_b(1-\nu_b)-D_a(1-\nu_a))}D_b.
\end{eqnarray}
\endnumparts
from wich we can obtain the effective refractive index of the cluster as
\beq
\label{eq:neff}
n_{eff}=\left(\frac{\rho_{eff}}{D_{eff}}\frac{D_b}{\rho_b}\right)^{1/4}.
\eeq

The effective refractive index is therefore a function of the physical properties of the inclusions and their filling fraction (i.e., their radius), so that changing these properties locally we can assume that we are creating an inhomogeneous material, that properly designed can create a gradient index device. This approach has been widely used with photonic and phononic crystals and here it is applied for flexural waves. 

The design method consist in defining a position dependent refractive index $n=n(\bm{r})$, then we have that the inclusion located at $\bm{r}_\alpha$ has associated a refractive index $n_{eff}=n(\bm{r}_\alpha)$. After that, equation \eref{eq:neff} is solved to determine the corresponding filling fraction, since $n_{eff}=n(f_\alpha)$, from which we obtain the radius of the inclusion at the given position.

We propose to study a family of circular refractive lenses, given their omnidirectional characteristics, where the refractive index is a function of the radial coordinate only, that is, $n(\bm{r})=n(r)$. The devices will be designed for a silicon plate of thickness $h_b=0.1a$, being $a$ the lattice constant of the arrangement of inclusions. The proposed lenses are the Luneburg and Maxwell lens, studied in section \ref{sec:LNMX} and the beam splitter, studied in section \ref{sec:BS}. The lens is made by means of the cluster of inclusions shown in the left panel of figure \ref{fig:schematics}, being the radius of the cluster $R_c=10.5a$ and where the radius of the inclusion at a given position will be selected according to the aforementioned method.

The design of efficient circular gradient index lenses requires a wide variation of the effective refractive index, variation that cannot be achieved by all type of inclusions. Right panel of figure \ref{fig:schematics} shows the effective refractive index for flexural waves in silicon as a function of the filling fraction for different materials' inclusions, whose elastic parameters are given in table \ref{tab:materials}. It is clear that lead inclusions give us the wider variation of the refractive index and for this reason they will be employed in the following calculations.

\begin{table}
\centering
\begin{tabular}{|c|c|c|c|c|}
\hline
 Parameter/Material    & Silicon & Lead & Copper & Nickel \\ \hline \hline
 $\rho\, (Kg/dm^3)$ & 2.329          &   11.34   & 8.96 & 8.91 \\ \hline
	$E\, (GPa)$ & 150      &   16   & 115 & 200 \\ \hline
  $\nu$ & 0.28          &  0.44    & 0.34 & 0.31 \\ \hline
\end{tabular}
\caption{\label{tab:materials} Elastic constants of the materials used for the simulations}
\end{table}

\section{Luneburg and Maxwell Lens}
\label{sec:LNMX}
In this section the realization and properties of the Luneburg and Maxwell lens will be studied (see for instance reference \cite{vsarbort2012spherical} for their detailed description). The Luneburg lens consists of a circular or spherical inhomogeneous lens such that the focal point is located at the lens boundary, so that a plane wave coming from the infinite always focus on the opposite border of the lens. The Maxwell lens consists of a similar device but here a point source at the border of the lens is always focused at the opposite border. These two lenses are inhomogeneous, being the refractive index as a function of the radial coordinate $r$ for the Luneburg lens given by
\beq
\label{eq:nLN}
n_L(r)=\sqrt{2-\frac{r^2}{R_c^2}},
\eeq
while for the Maxwell lens is given by
\beq
\label{eq:nMX}
n_M(r)=\frac{1}{1+(r/R_c)^2}.
\eeq

\begin{figure}\centering\centering
\includegraphics[scale=1]{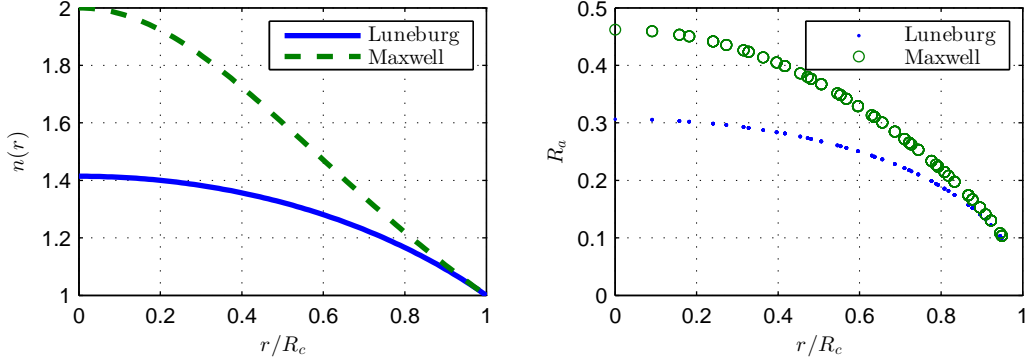}
\caption{\label{fig:neff_LNMX}
Left panel: Refractive index as a function of the radial coordinate for a Luneburg (blue continuous line) and a Maxwell (green dashed line) lens. Right panel: Radius of the lead inclusions as a function of the distance to the center for the circular cluster for the Luneburg (blue dots) and Maxwell (green circles) lens.
				}
\end{figure}

Figure \ref{fig:neff_LNMX}, left panel, shows the refractive index as a function of the distance to the center of the lens for the Luneburg lens (blue continuous line) and for the Maxwell lens (green dashed line). It is clear that the variation of the refractive index is higher for the Maxwell lens, since the ray trajectories, as will be seen later, have to follow a more curved path. We see in the figure and from equations \eref{eq:nLN} and \eref{eq:nMX} than the maximum refractive index required for the Luneburg lens is at $r=0$ and is $n_{max}=1.414$, while for the Maxwell lens is, as well at $r=0$, $n_{max}=2$. 

The right panel of Fig. \ref{fig:neff_LNMX} shows the variation of the inclusions' radii as a function of their distance to the center of the cluster for the realization of the Luneburg and Maxwell lens. As mentioned before, the plate's material is silicon and the inclusions are made of lead. Obviously, as we are closer to the center of the lens, the refractive index has to be higher, or the effective velocity lower, so that the radius of the inclusions must be higher in order to enhance the scattering effects, which slow-down the wave.

Multiple scattering simulations \cite{lee2010scattering} have been performed in order to check the validity of the design and visualize the behaviour of the lenses for different wavelengths. In all the simulations we have chosen four wavelengths, being $\lambda=3,45$ and $6$ times the lattice constant of the arrangements of inclusions. It is expected that the effective medium description be accurate for wavelengths larger than 4 times the lattice constant, while diffraction effects can hinder its functionality for very large wavelengths.

\begin{figure}\centering\centering
\includegraphics[scale=1]{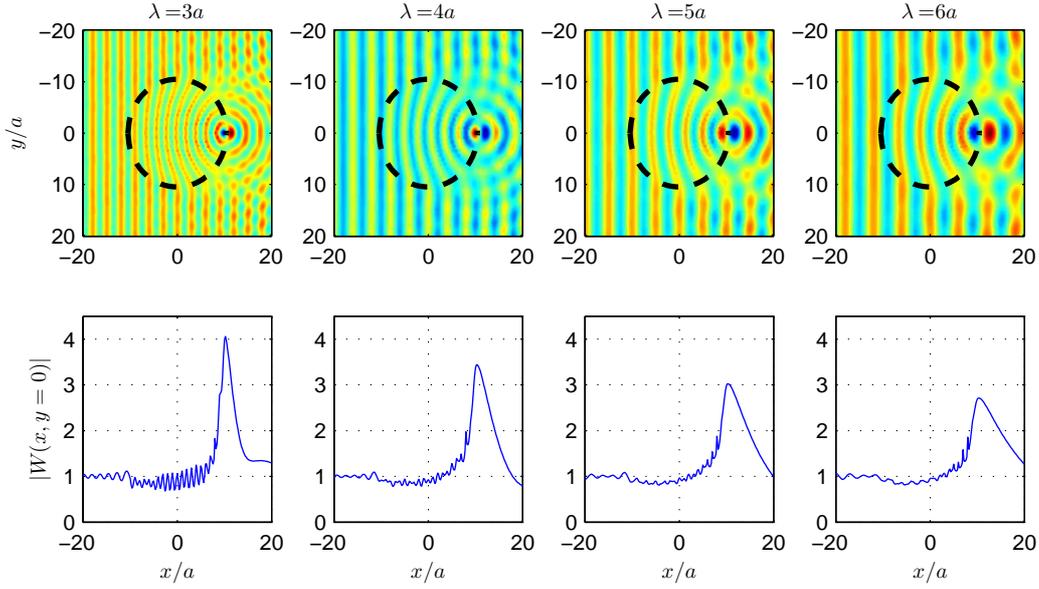}
\caption{\label{fig:mstLNPW}
Multiple scattering simulations of a circular cluster of lead inclusions whose radii have been defined
in order to create a Luneburg lens (see text for details). The position of the inclusions is not shown for clarity. The simulation consists in a plane wave arriving from the left and interacting with the cluster of inclusions, then a focusing point appears at the border of the right hand side of the cluster. The corresponding wavelengths are depicted in the title of each graph. The upper panel shows the field distributions along the $xy$ plane, while the lower panels shows the field profiles along the line $y=0$.}
\end{figure}

Figure \ref{fig:mstLNPW}, upper panels, shows the cluster of inclusions designed to behave like a Luneburg lens, and it shows how a plane wave comes from the left and impinges the cluster. The focusing point at the opposite border of the lens is evident for the four wavelengths studied, showing the good performance of this design. The lower panels shows the field profile along the horizontal axis of the lens ($y=0$), it is clear how a strong focusing point appears at the opposite border of the lens, being its amplitude between three and four times that of the incident plane wave.

\begin{figure}\centering\centering
\includegraphics[scale=1]{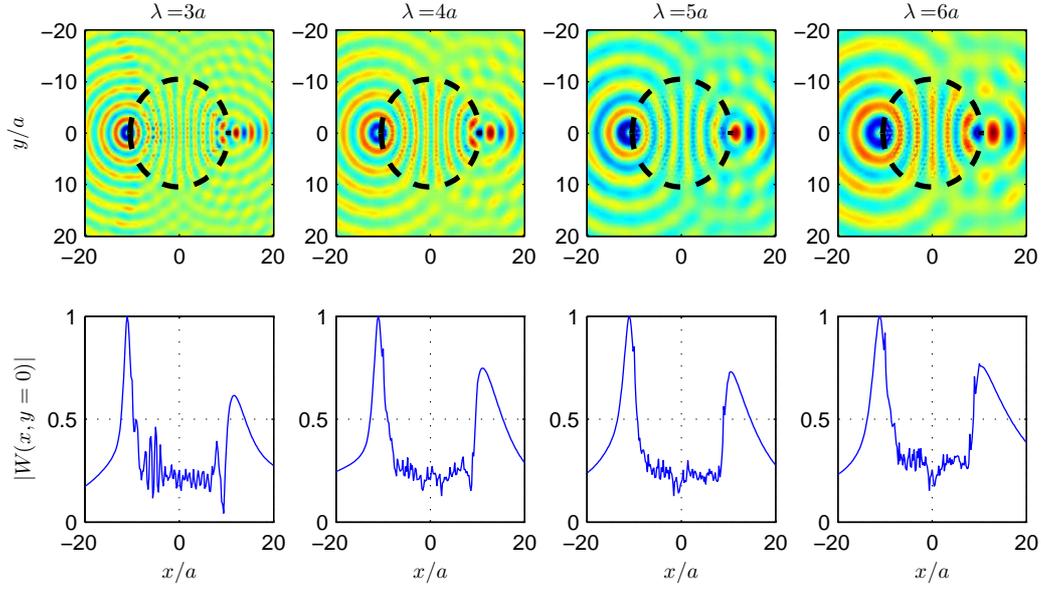}
\caption{\label{fig:mstMXPS}
Similar system of Fig. \ref{fig:mstLNPW} but with the radii of lead inclusions defined to create a Maxwell lens. A point source is located at $x=-Rc$ and a focusing point appears at $x=R_c$. 
				}
\end{figure}

Figure \ref{fig:mstMXPS}, upper panels, shows the cluster of inclusions designed to behave like a Maxwell lens. In this case, a point source is located at $x=-R_c$ and its image is reconstructed at $x=R_c$, and this behaviour is maintained for the four wavelengths considered. Lower panels show the field profile along the axis of the lens, it is clear that the focusing point is wider than the source, given that the evanescent components of the point source do not propagate through the lens. Also, we see that the amplitude of the field at the focusing point is smaller than at source.

\begin{figure}\centering\centering
\includegraphics[scale=1]{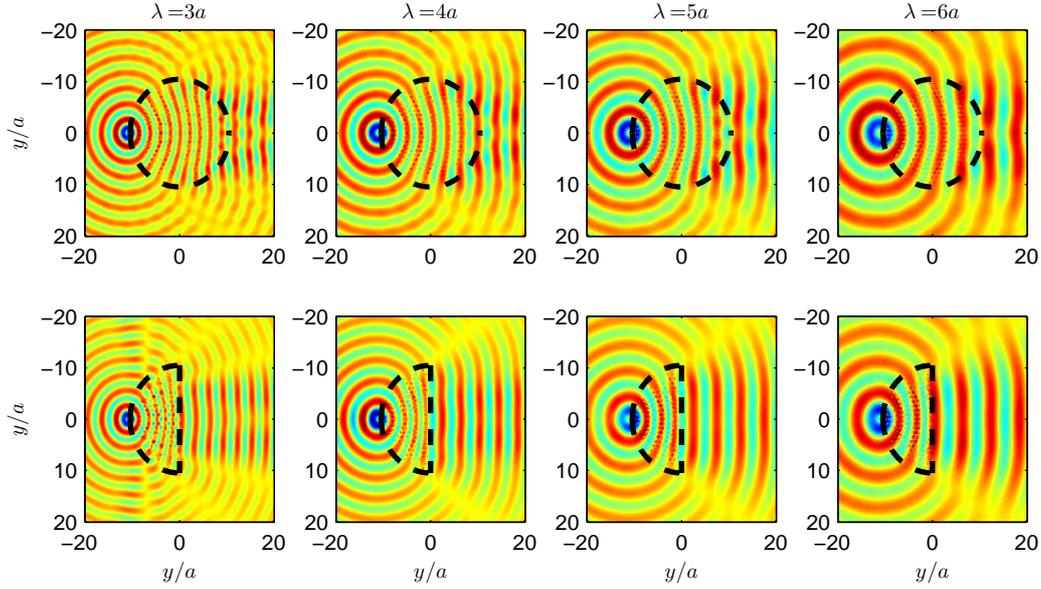}
\caption{\label{fig:mstPWgenerators}
Illustration of the behaviour of the Luneburg (upper panels) and Maxwell (lower panels) lenses as plane-wave generators. For the Maxwell lens only one half of the cluster of Fig. \ref{fig:mstMXPS} is employed so that the wave front is parallel to the lens' surface, then a plane wave is generated.}
\end{figure}

Finally, figure \ref{fig:mstPWgenerators} shows the previous lenses working as plane wave generators from point sources, the upper panels showing the Luneburg lens and the lower panels showing the Maxwell lens. In this last case only one half of the lens is implemented, given that it is at the middle of the lens where the wave-front becomes flat. Then, a point source is located at the border of the lens and it is focused at infinity, that is, it is converted to a nearly  plane wave. It is clear that the performance of the lens is better for wavelengths of $\lambda=3a$ or $\lambda=4a$, since larger wavelengths, given the size of the cluster, cannot define very well a plane wave-front due to diffraction effects. Shorter wavelengths would be better for the definition of the plane wave but the effective medium theory is not valid for wavelengths shorter than 4 or 3 times the lattice constant $a$. The design based on the Maxwell lens has the advantage, from the practical point of view, of requiring only one half of the scatterers than the Luneburg lens, then it is indeed a smaller device. Also, the quality of the plane wave-front looks better than those shown for the Luneburg lens. However, the omni-directionality nature of the device is obviously lost here.

\section{Beam Splitters}
\label{sec:BS}
In this section a special family of refractive circular lenses is studied, called ``beam splitters''. In these devices the profile of the refractive index
is defined in such a way that a plane wave arriving to them is divided in two beams, each one propagating with deflection angles $\theta_0$ and $-\theta_0$ with respect the initial propagation direction. 

The exact profile for this type of lenses is obtained from the solution of the following equation \cite{vsarbort2012spherical}
\beq
\label{eq:neff_BS}
(r/R_c)-2n^{\alpha-1}+(r/R_c)n^{2\alpha}=0,
\eeq
where $\alpha$ is related by the deflection angle as $\theta_0=\pi/\alpha$. 

\begin{figure}\centering\centering
\includegraphics[scale=1]{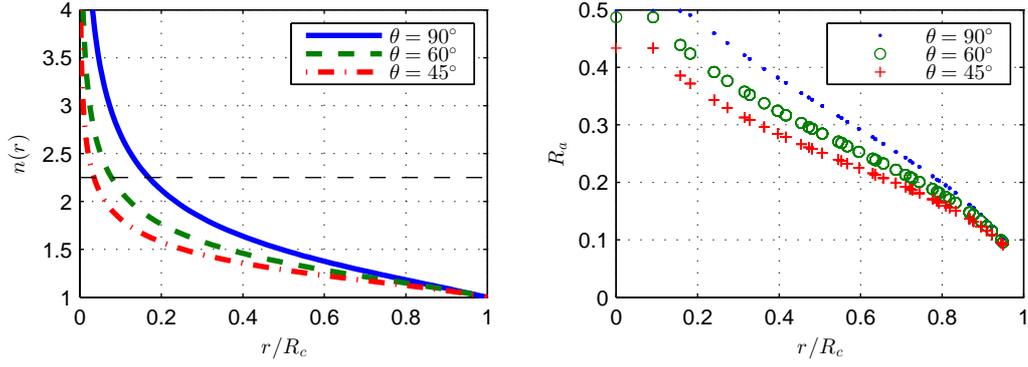}
\caption{\label{fig:neff_BS}
Left panel: Effective refractive index as a function of the radial coordinate for the ``beam splitter'' described in the text. Results are shown for deflection angles of 90 (blue line), 60 (green line) and 45 (red line) degrees. Right panel: Radius of the lead inclusions in the plate as a function of the distance to the center of the cluster. 
}
\end{figure}

Figure \ref{fig:neff_BS}, left panel, shows the solution for $n(r)$ for three different
deflection angles, $\theta_0=90^\circ,60^\circ$ and $45^\circ$. It is obvious that as the deflection angle increases, the variation of the refractive index is higher, and this variation is not always possible. The horizontal line in the figure shows the maximum refractive index that is achievable by means of lead inclusions, inclusions below this line will not have the proper size. It means that the fabricated lens is not ``perfect'' but they still have the proper functionality, as will be seen later. The right panel shows the corresponding radius of the inclusions as a function of their distance to the center of the cluster.

\begin{figure}\centering\centering
\includegraphics[scale=1]{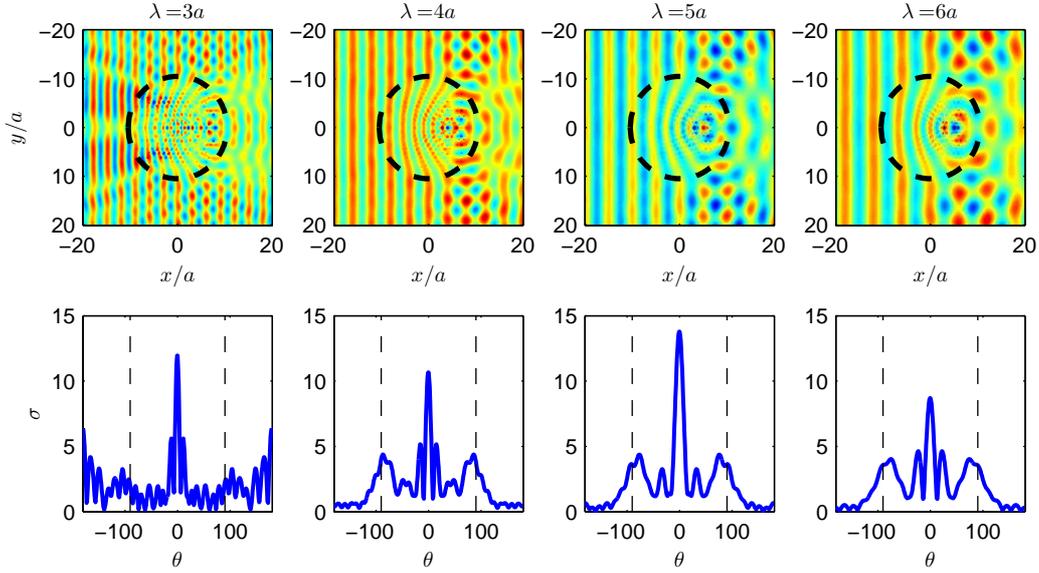}
\caption{\label{fig:mstBSalpha2}
Upper panels: Multiple scattering simulation at different wavelengths showing the behaviour of the cluster as a beam
splitter, being the deflection angle of $90^\circ$. Lower panels: Far field amplitude as a function of the polar angle $\theta$.
It is seen that for a wavelength $\lambda=3a$ the splitter is not working properly, but for larger wavelengths an interference pattern 
can be seen in the perpendicular direction.				}
\end{figure}

Like in the preceding section, multiple scattering simulations have been performed in order to visualize the behaviour of the lenses for finite wavelengths. Then, Fig. \ref{fig:mstBSalpha2}, upper panels, shows the behaviour of the cluster acting as a beam splitter of $\theta_0=90^\circ$, that is, $\alpha=2$ in equation \eref{eq:neff_BS}. We can see that for a wavelength $\lambda=3a$ the beam splitter is not working properly, while it is clear that there is a bending of the plane wave inside the cluster, the output wave is not in the right direction. The situation changes for the other wavelengths, where the interference patterns clearly show a propagating wave in the perpendicular direction. 

The lower panels of Fig. \ref{fig:mstBSalpha2} show the far field pattern as a function of the polar angle $\theta$. We can see now that, except for $\lambda=3a$, peaks in this pattern appear at the proper directions of $\theta=\pm 90^\circ$, represented by the vertical dashed lines. The central peak corresponding to the far field amplitude in the forward direction appears due to the optical theorem for flexural waves, which relates the total scattering cross section with the amplitude in the forward direction \cite{NorrisVemula}, and it is a consequence of energy conservation.

\begin{figure}\centering\centering
\includegraphics[scale=1]{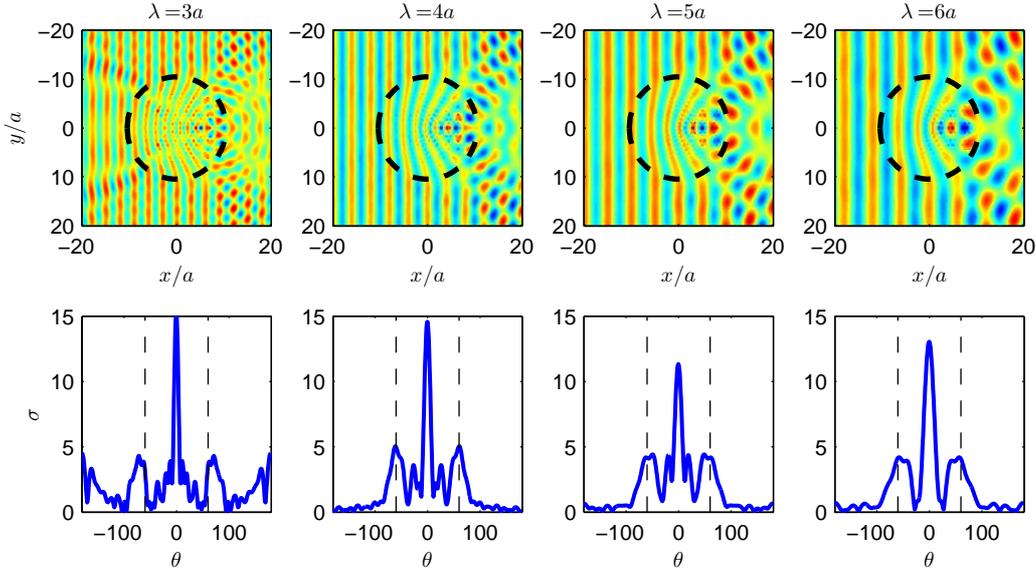}
\caption{\label{fig:mstBSalpha3}
Similar system as in Fig.\ref{fig:mstBSalpha2}, but changing the deflection angle to $60^\circ$. The splitter is working properly at all wavelengths in this case.
				}
\end{figure}

Figure \ref{fig:mstBSalpha3}, upper panels, shows the cluster acting as a beam splitter with $\theta_0=60^\circ$, in this case the interference pattern clearly shows that the beam splitter is working properly for the four wavelengths, and the far field amplitude shown in the lower panels support this fact.

\begin{figure}\centering
\includegraphics[scale=1]{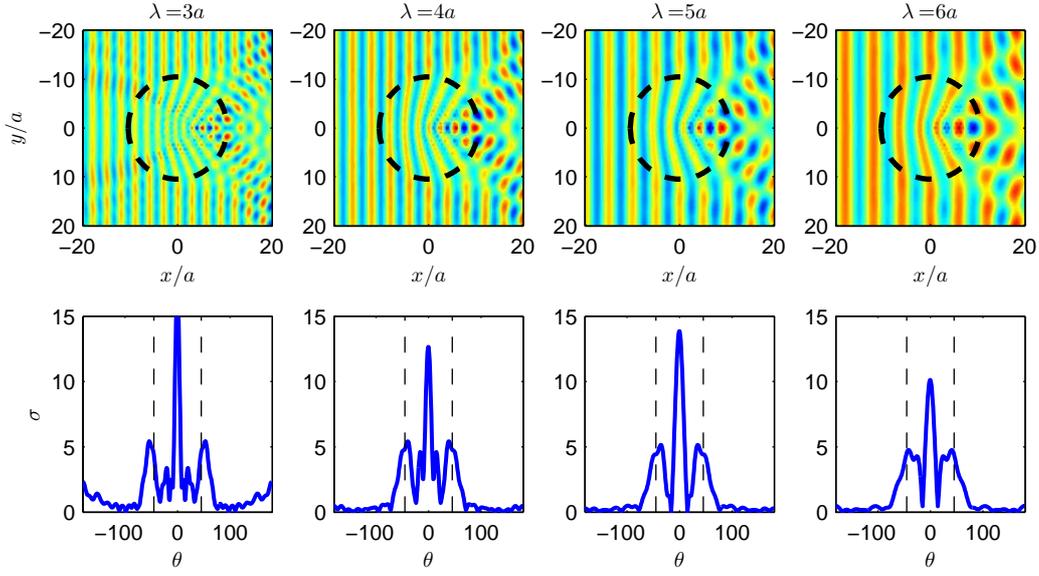}
\caption{\label{fig:mstBSalpha4}
Similar system as in Fig.\ref{fig:mstBSalpha2}, but changing the deflection angle to $45^\circ$. The splitter is working properly at all wavelengths in this case.				}
\end{figure}

Similarly, figure \ref{fig:mstBSalpha4}, upper panels, shows multiple scattering simulations for the cluster designed as a beam splitter with $\theta_0=45^\circ$, and the good performance of the splitter can be observed from the interference pattern and the far field amplitude shown in the lower panels.

Although full wave simulations can visualize the direction of the deflected beam, the analysis of the far field can give quantitative information about this direction. The lower panels of figures \ref{fig:mstBSalpha2},\ref{fig:mstBSalpha3} and \ref{fig:mstBSalpha4} shows that the far field patterns for the three beam splitters present strong peaks at the expected directions, although a frequency response of these peaks is obviously expected. Thus, for short wavelengths the splitters will not work, given that effective medium description presents a cut-off wavelength below which the field detects the inhomogeneous nature of the phononic crystal, and for longer wavelengths diffraction effects will hinder the device's functionality.

\begin{figure}
\centering
\includegraphics[scale=1]{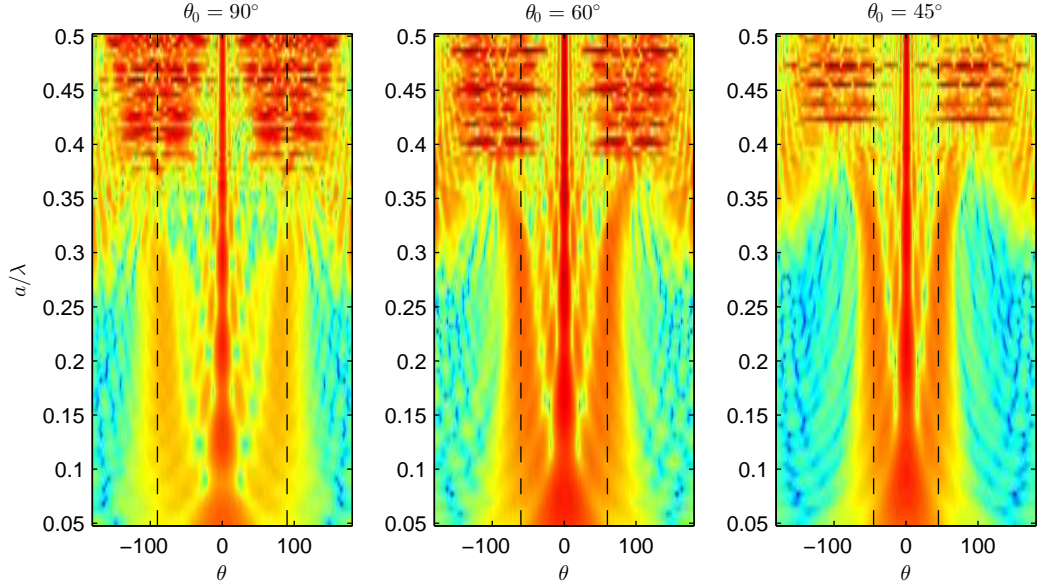}
\caption{\label{fig:sigmaBSalpha234}
Far field pattern corresponding to the $\theta_0=60^\circ$ beam splitter as a function of the reduced frequency $a/\lambda$. Vertical lines shows the directions at which peaks are expected, and we see that they appear for a wide frequency region.
				}
\end{figure}

In order to quantitatively define these upper and lower limits, we have analyzed the far field pattern as a function of frequency for the three beam splitters. Then, figure \ref{fig:sigmaBSalpha234} shows the far field amplitude for the three beam splitters previously considered as a function of the reduced frequency $a/\lambda$. The expected behaviour described before for these devices is clear in the figure. For very low frequencies the far field is focused in the forward direction, with weak peaks at the expected deflection angles, which correspond to the vertical dashed lines. When the wavelength is of the order of the cluster's size $R_c=10.5a$, we see how the peaks are important where expected, and the beam splitter is working properly, but as we go beyond the homogenization limit $a/\lambda>0.25a$, we see how the peaks deviate from the expected angles, given to the fact that the effective medium description is not a accurate and the propagation velocity is lower than expected, making that the cluster behave as a beam splitter with a higher $\theta_0$. Finally, above $a/\lambda=0.4$, a band gap in the structure is expected and  Bragg reflection occurs, so that we can see that the cluster is highly dispersive.

Therefore, although beam splitters have very demanding refractive index variations, specially as the deflection angle increases, ``imperfect'' realizations of them, where the radius of the inclusions at the center of the device might not be correct, work properly in the homogenization region ($\lambda\geq 4a$).


\section{Summary}
\label{sec:summary}
In summary, a set of omnidirectional devices for flexural waves based on phononic crystals have been
presented, and a very good performance in a wide frequency region has been demonstrated.

The devices studied consisted in the Luneburg and Maxwell lenses and beam splitters, which are circular inhomogeneous regions where the refractive index is a specific function of the radial coordinate, working then as an omnidirectional gradient index devices.

Their realization has been studied by means of lead inclusions in a silicon matrix, where a circular cluster of these inclusions has been created. This cluster behaves, in the low frequency limit, as an homogeneous material with a specific refractive index, obtained by means of a homogenization theory that relates the effective parameters of the cluster with the physical properties and size of the inclusions. The required variation of the refractive index is therefore obtained by the corresponding variation of the inclusions' radii.

Multiple scattering simulations have been performed to verify the functionality of the proposed devices, and also to visualize their possible applications as focusing or beam forming devices. It is found that they work as expected in a wide frequency region, limited by the so-called homogenization limit, corresponding to wavelengths larger than 3 or 4 times the distance between inclusions. 

This work offers an alternate way for the realization of these devices than those previously presented by means of thickness variations, and it is specially useful at the nano-scale, with potential applications for beam forming and energy harvesting.

\section*{Acknowledgments}
Work supported by the Agence Nationale de la Recherche and Direction G\'en\'erale de l'Armement under the project Metactif, grant ANR-11-ASTR-015.
\section*{References}

\begin{thebibliography}{10}

\bibitem{Evans2007}
D.~V. Evans and R.~Porter.
\newblock {Penetration of flexural waves through a periodically constrained
  thin elastic plate in vacuo and floating on water}.
\newblock {\em Journal of Engineering Mathematics}, 58(1-4):317--337, March
  2007.

\bibitem{mcphedran2009platonic}
RC~McPhedran, AB~Movchan, and NV~Movchan.
\newblock Platonic crystals: Bloch bands, neutrality and defects.
\newblock {\em Mechanics of Materials}, 41(4):356--363, 2009.

\bibitem{Movchan2007}
A~B Movchan, N~V Movchan, and R~C McPhedran.
\newblock {Bloch-Floquet bending waves in perforated thin plates}.
\newblock {\em Proceedings of the Royal Society A: Mathematical, Physical and
  Engineering Sciences}, 463(2086):2505--2518, October 2007.

\bibitem{pennec2008low}
Y~Pennec, B~Djafari-Rouhani, H~Larabi, JO~Vasseur, and AC~Hladky-Hennion.
\newblock Low-frequency gaps in a phononic crystal constituted of cylindrical
  dots deposited on a thin homogeneous plate.
\newblock {\em Physical Review B}, 78(10):104105, 2008.

\bibitem{pennec2009phonon}
Y~Pennec, B~Djafari Rouhani, H~Larabi, A~Akjouj, JN~Gillet, JO~Vasseur, and
  G~Thabet.
\newblock Phonon transport and waveguiding in a phononic crystal made up of
  cylindrical dots on a thin homogeneous plate.
\newblock {\em Physical Review B}, 80(14):144302, 2009.

\bibitem{marchal2012dynamics}
R~Marchal, O~Boyko, B~Bonello, J~Zhao, L~Belliard, M~Oudich, Y~Pennec, and
  B~Djafari-Rouhani.
\newblock Dynamics of confined cavity modes in a phononic crystal slab
  investigated by in situ time-resolved experiments.
\newblock {\em Physical Review B}, 86(22):224302, 2012.

\bibitem{Xiao2011}
Yong Xiao, Brian~R. Mace, Jihong Wen, and Xisen Wen.
\newblock {Formation and coupling of band gaps in a locally resonant elastic
  system comprising a string with attached resonators}.
\newblock {\em Physics Letters A}, 375(12):1485--1491, March 2011.

\bibitem{Xiao2012}
Yong Xiao, Jihong Wen, and Xisen Wen.
\newblock {Flexural wave band gaps in locally resonant thin plates with
  periodically attached spring-mass resonators}.
\newblock {\em Journal of Physics D: Applied Physics}, 45(19):195401, May 2012.

\bibitem{Torrent2013Graphene}
Daniel Torrent, Didier Mayou, and Jose Sanchez-Dehesa.
\newblock {Elastic analog of graphene: Dirac cones and edge states for flexural
  waves in thin plates}.
\newblock {\em {PHYSICAL REVIEW B}}, {87}({11}), {MAR 27} {2013}.

\bibitem{Zhu2012}
R.~Zhu, X.~N. Liu, G.~L. Huang, H.~H. Huang, and C.~T. Sun.
\newblock {Microstructural design and experimental validation of elastic
  metamaterial plates with anisotropic mass density}.
\newblock {\em Physical Review B}, 86(14):144307, October 2012.

\bibitem{Farhat2009}
Mohamed Farhat, Sebastien Guenneau, Stefan Enoch, and Alexander Movchan.
\newblock {Cloaking bending waves propagating in thin elastic plates}.
\newblock {\em Physical Review B}, 79(3):033102, January 2009.

\bibitem{Farhat2009a}
Mohamed Farhat, Sebastien Guenneau, and Stefan Enoch.
\newblock {Ultrabroadband Elastic Cloaking in Thin Plates}.
\newblock {\em Physical Review Letters}, 103(2):1--4, July 2009.

\bibitem{Stenger2012}
Nicolas Stenger, Manfred Wilhelm, and Martin Wegener.
\newblock {Experiments on Elastic Cloaking in Thin Plates}.
\newblock {\em Physical Review Letters}, 108(1):1--5, January 2012.

\bibitem{Farhat2010}
Mohamed Farhat, Sebastien Guenneau, Stefan Enoch, Alexander~B. Movchan, and
  Gunnar~G. Petursson.
\newblock {Focusing bending waves via negative refraction in perforated thin
  plates}.
\newblock {\em Applied Physics Letters}, 96(8):081909, 2010.

\bibitem{pierre2010negative}
J~Pierre, O~Boyko, L~Belliard, JO~Vasseur, and B~Bonello.
\newblock Negative refraction of zero order flexural lamb waves through a
  two-dimensional phononic crystal.
\newblock {\em Applied Physics Letters}, 97(12):121919--121919, 2010.

\bibitem{lenteTTWu}
Tsung-Tsong Wu, Yan-Ting Chen, Jia-Hong Sun, Sz-Chin~Steven Lin, and Tony~Jun
  Huang.
\newblock Focusing of the lowest antisymmetric lamb wave in a gradient-index
  phononic crystal plate.
\newblock {\em App. Phys. Lett.}, 98(17):171911, 2011.

\bibitem{zhao2012efficient}
Jinfeng Zhao, R{\'e}mi Marchal, Bernard Bonello, and Olga Boyko.
\newblock Efficient focalization of antisymmetric lamb waves in gradient-index
  phononic crystal plates.
\newblock {\em Applied Physics Letters}, 101(26):261905, 2012.

\bibitem{krylov2012resumen}
V.V. Krylov.
\newblock Acoustic black holes and their applications for vibration damping and
  sound absorption.
\newblock In {\em Proceedings of the International Conference on Noise and
  Vibration Engineering (ISMA 2012)}, pages 933--944. Sas, P., Moens, D. and
  Jonckheer, S. (eds.)., Sept 2012.

\bibitem{climente2013omnidirectional}
Alfonso Climente, Daniel Torrent, and Jos{\'e} S{\'a}nchez-Dehesa.
\newblock Omnidirectional broadband insulating device for flexural waves in
  thin plates.
\newblock {\em Journal of Applied Physics}, 114(21):214903, 2013.

\bibitem{climente2014gradient}
Alfonso Climente, Daniel Torrent, and Jos{\'e} S{\'a}nchez-Dehesa.
\newblock Gradient index lenses for flexural waves based on thickness
  variations.
\newblock {\em Applied Physics Letters}, 105(6):064101, 2014.

\bibitem{torrent2014effective}
Daniel Torrent, Yan Pennec, and Bahram Djafari-Rouhani.
\newblock Effective medium theory for elastic metamaterials in thin elastic
  plates.
\newblock {\em Phys. Rev. B}, 90:104110, Sep 2014.

\bibitem{Graff}
K.~F. Graff.
\newblock {\em Wave Motion in elastic solids, 2nd Ed}.
\newblock Dover, 1991.

\bibitem{vsarbort2012spherical}
Martin {\v{S}}arbort and Tom{\'a}{\v{s}} Tyc.
\newblock Spherical media and geodesic lenses in geometrical optics.
\newblock {\em Journal of Optics}, 14(7):075705, 2012.

\bibitem{lee2010scattering}
Wei-Ming Lee and Jeng-Tzong Chen.
\newblock Scattering of flexural wave in a thin plate with multiple circular
  holes by using the multipole trefftz method.
\newblock {\em International Journal of Solids and Structures},
  47(9):1118--1129, 2010.

\bibitem{NorrisVemula}
A.N. Norris and C.~Vemula.
\newblock Scattering of flexural waves on thin plates.
\newblock {\em Journal of Sound and Vibration}, 181(1):115 -- 125, 1995.

\end{thebibliography}

\end{document}